\documentclass[11pt,twoside]{article}

\usepackage{asp2006}
\usepackage{epsf}
\usepackage{psfig}
\usepackage{lscape}
\usepackage{graphicx}

\begin{document}
\title{Searching for Hyper-Velocity Stars}   
\author{A. Tillich, U. Heber and H. Hirsch}
\affil{Dr.Remeis-Sternwarte Bamberg, Universit\"at Erlangen-N\"urnberg, 
Sternwartstr.7, D-96049 Bamberg, Germany}

\begin{abstract}
We present our survey for subluminous hyper-velocity candidates, which has been successfully
initiated at the ESO NTT and the Calar Alto 3.5\,m telescope.
\end{abstract}

A natural consequence of the presence of a super-massive black hole in the 
centre of
our Galaxy is the existence of dynamically ejected stars, called 
hyper-velocity stars (HVS).
Up to now ten HVSs are known that have Galactic rest-frame velocities 
exceeding the escape velocity of the Galaxy and are therefore unbound. 
To date only one of those, the
subluminous O star US\,708 (Hirsch et al. 2005), belongs to an old population
 of evolved stars. 
Brown et al. (2007)
 found 26 bound late-B-type HVSs 
 with $v_{\mathrm GRF}\,>\,275\,\mathrm{km\,s^{-1}}$ but only one bound HVS
 with $v_{\mathrm GRF}\,<\,-275\,\mathrm{km\,s^{-1}}$ supporting the
 perception that late-B-type HVSs are short-lived main sequence rather than
 long-lived HB stars.
If such a bound population of hot subdwarf stars, no asymmetry in the radial
velocity distribution is expected to occur because these stars are long lived.
Hence there should also be a population of subdwarf stars moving towards us
at high speed.  

Therefore we embarked on a search for new HVSs among 
subdwarf O and B stars (unbound or bound).
An excellent starting point is 
the enormous data base
of the Sloan Digital Sky Survey (SDSS). By means of spectral classification a 
number of sdOB HVS candidates has been extracted, which we follow-up presently.

 

In order to rule out radial velocity variability, hence an indication 
for binarity, we need to take two spectra.
By determining the atmospheric parameters we 
estimate the spectroscopic distance.
Calculating possible trajectories from the GC we derive information about 
the ejection velocity, the time of flight and can predict the corresponding 
proper motions, which are expected to be so small that they can probably be 
verified 
only by means of the GAIA satellite 
mission.

Our first observing run was scheduled for February 2007 
at the Calar Alto 3.5\,m telescope and was clouded out completely. 
Observations at the ESO NTT have recently finished and are currently analysed. 
The survey will be continued by observations at the Calar Alto observatory and 
the ESO VLT in spring/summer of 2008.


\end{document}